\documentstyle[aps,prb,twocolumn,floats,psfig]{revtex}

\begin{document}

\bibliographystyle{prsty}

\title{
Decoherence of a Superposition of Macroscopic Current States in a
SQUID
\vspace{-1mm} }

\author{
E. M. Chudnovsky$^1$ and A. B. Kuklov$^2$}

\address{ $^1$ Department of Physics and Astronomy, Lehman College,
City University of
New York, \\
250 Bedford Park Boulevard West, Bronx, New York 10468-1589
\\
$^2$ Department of Engineering Science and Physics, The College of
Staten Island, \\
City University of New York, Staten Island, New York 10314
\smallskip
\bigskip\\
\parbox{14.2cm}
{\rm We show that fundamental conservation laws mandate
parameter-free mechanisms of decoherence of quantum oscillations
of the superconducting current between opposite directions in a
SQUID --- emission of phonons and photons at the oscillation
frequency. The corresponding rates are computed and compared with
experimental findings. The decohering effects of external
mechanical and magnetic noise are investigated.
\smallskip
\begin{flushleft}
\end{flushleft}
}}
\maketitle

\section{Introduction}

The possibility of macroscopic quantum tunneling of the magnetic
flux in SQUIDs was suggested in a seminal paper of Caldeira and
Leggett \cite{Cal-Leg} and was subsequently demonstrated in
experiment \cite{Martinis,Clarke}. Following these developments
the possibility of coherent quantum oscillations between
macroscopic flux states was intensively studied by theorists (see,
e.g. Ref.\ \onlinecite{Leggett} and references therein). Modern
interest to this problem was generated by the hope to build a
superconducting (SC) qubit. The goal of preparing quantum
superposition of macroscopic flux states in a SQUID had remained
elusive, however, until recent experiments of Friedman et al.
\cite{Friedman} and van der Wal et al. \cite{Mooij}. In these
experiments the tunneling splitting between the states
corresponding to symmetric and antisymmetric quantum
superpositions of clockwise and counterclockwise currents in a
SQUID loop have been observed. These experiments were followed by
similar measurements of more elaborate designs \cite{Vion,Yu}.

The most important question about the above experiments is the one
of the decoherence time \cite{Leggett-Science}. Any degree of
freedom that interacts with the coherently oscillating variable
can be the source of decoherence. One can divide all such sources
into two groups, avoidable and unavoidable. Examples of avoidable
sources are, e.g., nuclear spins and non-thermal noise. In this
paper we study generic mechanisms of decoherence which are
controlled by the conservation laws \cite{Chudnovsky,Chu-Mar}. In
application to SQUIDs, such mechanisms are unavoidable as they
originate from the fundamental symmetries of free space.  To
illustrate our point, consider, e.g., experiment of Ref.\
\onlinecite{Friedman} in which the current,
$J\,{\sim}\,3\,{\mu}$A, oscillated due to quantum tunneling
between clockwise and counterclockwise directions. The angular
momentum associated with the current was of order
$L_{c}\,{\sim}\,10^{10}{\hbar}$. The total angular momentum of the
system is $L_{tot}=L_{c}+L_{m}$, where $L_m$ is the mechanical
angular momentum of the solid matrix bearing the current. To
conserve $L_{tot}$, oscillations of $\;L_{c}$ between
$L_{c}=10^{10}{\hbar}$ and $L_{c}=-10^{10}{\hbar}$ must be
accompanied by simultaneous torsional oscillations of the solid
matrix between the states with the angular momenta
$L_{m}=-10^{10}{\hbar}$ and $L_{m}=10^{10}{\hbar}$, so that
$L_{tot}=0$.  One can argue that the current loop is always firmly
attached to a substrate which is firmly attached to another solid,
etc., so that the angular momentum is transferred to the infinite
mass, like the linear momentum in M\"{o}ssbauer experiment. If
this were true, the SC current would simply bounce elastically
between clockwise and counterclockwise directions, making
conservation of the total angular momentum irrelevant. It is easy
to see, however, that for a current oscillating at a high
frequency, e.g., $f_{0}\,{\sim}\,10^{9}-10^{10}\,$s$^{-1}$, the
analogy with the M\"{o}ssbauer effect breaks down. Indeed, during
the period of oscillations, the elastic deformation cannot travel
more than a distance $v_{s}/f_{0}$ away from the current; $v_{s}$
being the speed of the transverse sound. For, e.g.,
$v_{s}\,{\sim}\,5{\times}10^{3}\,$m/s and
$f_{0}\,{\sim}\,5{\times}10^{9}\,$s$^{-1}$ one obtains
$v_{s}/f_{0}\,{\sim}\,1\,{\mu}$m. Thus, the part of the solid
involved in the conservation of the angular momentum is small, not
macroscopically large. Consequently, the torsional oscillations of
the part of the solid matrix ``co-wiggling" with the current must
generate phonons of frequency $f_{0}$ in the surrounding matter.
As long as the speed of sound is finite this should result in a
decoherence of quantum oscillations of the current. Similar effect
exists due to the interaction of the magnetic moment of the
current with vacuum photons. The difference from the phonon
problem is that at $f_{0}\,{\sim}\,5{\times}10^{9}\,$s$^{-1}$ the
wavelength of the light, ${\lambda}_{l}=c/f_{0}$, is about five
orders of magnitude larger than ~$\lambda_s$. Consequently, the
vacuum properties of the photons depend strongly on the geometry
of the experiment, in particular, on the metal shielding of the
SQUID.

The above picture is quasiclassical. From the quantum mechanical
point of view, the states of the tunneling SC current are not
classified by specific values of the angular momentum. The
interaction of the current with the solid matrix entangles the
angular momentum states of the current with the angular momentum
states of the matrix. In the absence of the external noise, the
total angular momentum is, of course, a well-defined conserved
quantum number. In this respect, it is important to understand why
the external noise acting on the macroscopic solid matrix (like,
e.g., vibration of the building) does not instantaneously destroy
the quantum entanglement between the angular momentum states of
the current and the angular momentum states of the matrix. We will
discuss this issue in the context of the local versus global
mechanisms of decoherence.

The paper has the following structure. The decoherence due to the
exchange of the angular momentum with the solid matrix is studied
in Section II. The limits in which the dimensions of the SQUID
loop are smaller and greater than the wavelength of the emitted
phonons, ${\lambda}_{s}=v_{s}/f_{0}$, are considered in the
Subsections II-A and II-B correspondingly. The role of the
external mechanical noise is discussed in the Subsection II-C. The
photon effects are studied in Section III. The decoherence in the
unshielded and shielded SQUID are computed in Subsections III-A
and III-B, respectively. Conclusions that may be useful for
building SQUID-based qubits are summarized in Section IV.

\section{Decoherence due to phonons}

Consider quantum oscillations of the current $J$ in a flat loop of
spacial dimension $R$ at a frequency $f_{0}$. The borderline
between the cases of small and large SQUID studied in this Section
is the dimension $R\sim {\lambda}_{s}$. For $f_{0}$ of the order
of a few GHz, it is $R\,{\sim}\,1\,{\mu}$m. Obviously, both cases,
$R\,<\,{\lambda}_{s}$ and $R\,>\,{\lambda}_{s}$ must be of
interest for the ongoing experiments on quantum superposition of
SQUID states and in connection with the goal of building a SC
qubit.

Let the magnitude of the dimensionless angular momentum associated
with the current be $L$. The magnetic moment of the current is
\begin{equation}\label{M-L}
M={\mu}_{B}L=Ja/c\;,
\end{equation}
where ${\mu}_{B}$ is the Bohr magneton and $a$ is the area of the
loop. Thus, for a given current and given dimensions of the SQUID
the value of the oscillating angular momentum of the SC electrons
can be found from $L=Ja/{\mu}_{B}c$. For, e.g., a $0.1\,{\mu}$m
(small) SQUID carrying an electric current of
$J\,{\sim}\,0.1\,{\mu}$A, this gives $L\,{\sim}\,10^{2}$, while
for a $100\,{\mu}$m (large) SQUID carrying an electric current of
$J\,{\sim}\,10\,{\mu}$A, one obtains $L\,{\sim}\,10^{10}\,$.

The dynamical torsional deformations of the lattice are described
by the transversal displacement field ${\bf u}({\bf r}, t)$
satisfying
\begin{equation}\label{transversal}
{\bf \nabla}\cdot{\bf u}=0\;.
\end{equation}
These deformations do not affect the density of the ionic lattice
and, in the long-wave limit, do not modify electronic states. The
minimal, allowed by symmetry, coupling of such deformations with
the electric current ${\bf j}({\bf r}, t)$ must be proportional to
$ \int d^3r {\bf j}{\cdot}\dot{{\bf u}}$. Let us show that the
coefficient of proportionality is ${\rm m}_{e}/{\rm e}$, where
${\rm m}_{e}$ and ${\rm e}$ are bare electron mass and charge.
Without loss of generality, one can employ the following argument
in order to derive the interaction term. Consider the classical
density of the kinetic energy of electrons and ions,
\begin{equation}\label{KE}
{\rm KE} = n_{e}\frac{{\rm m}_{e}{\bf v}^2_e}{2} +
n_{i}\frac{M_{i}\dot{\bf u}^2}{2}\;,
\end{equation}
where $n_{e,i}$ are concentrations of electrons and ions, $M_{i}$
is the ionic mass, and ${\bf v}_{e}$ is the velocity field of
electrons in the laboratory coordinate frame. Notice now that the
electronic current is formed by the electron band states in the
coordinate frame co-moving with the lattice. Consequently,
\begin{equation}\label{def-j}
{\bf j}={\rm e}n_{e}({\bf v}_{e}-\dot{\bf u})\;.
\end{equation}
Expressing Eq.\ (\ref{KE}) in terms of physical variables ${\bf
j}$ and ${\bf u}$, one ends up with the coupling of the form
$({\rm m}_{e}/{\rm e})({\bf j}{\cdot}\dot{\bf u})$. The
generalization of this argument for quantum operators is trivial.
It gives the following Hamiltonian of the effective interaction in
the laboratory frame:
\begin{equation}\label{interaction}
{\cal H}_{int}= \frac{{\rm m}_{e}}{\rm e}\int d^3r\, {\bf
j}{\cdot}\dot{\bf u}\;.
\end{equation}
This formula is a consequence of the translational symmetry and
is, therefore, parameter-free. It can be used as long as ${\cal
H}_{int}$ results in a small perturbation of one-electron energy
band.

\subsection{Small SQUID}

We shall start with the case of a small current loop of
$R\,<\,{\lambda}_{s}$, in the $XY$ plane. The shape
of the loop will be irrelevant. In the context of its
magnetoelastic interaction with the solid matrix it is equivalent
to a magnetic atom of angular momentum $L$, imbedded in a solid.
We shall denote the states with ${\bf L}$ up and ${\bf L}$ down by
$|\uparrow\,\rangle$ and $|\downarrow\,\rangle$ respectively. They
are the eigenvalues of $L_{z}$:
\begin{equation}\label{L_z}
L_{z}|\uparrow\,\rangle=L|\uparrow\,\rangle\;,\;\;\;
L_{z}|\downarrow\,\rangle=-L|\downarrow\,\rangle\;.
\end{equation}
With no regard for the conservation of the angular momentum, the
ground state of the system can be approximated by
\begin{equation}\label{ground}
|0\,\rangle=\frac{1}{\sqrt{2}}(|\uparrow\,\rangle +
|\downarrow\,\rangle)\;,
\end{equation}
while the first excited state, with the same accuracy, is
\begin{equation}\label{first}
|1\,\rangle=\frac{1}{\sqrt{2}}(|\uparrow\,\rangle -
|\downarrow\,\rangle)\;.
\end{equation}
Let the energy separation of these two states be
$\Delta={\hbar}{\omega}_{0}=hf_{0}$. If the energy scale of the
experiment does not significantly exceed ${\Delta}$, then any
quantum state of the SQUID is a superposition
$|\psi\,\rangle=C_{1}|0\,\rangle + C_{2}|1\,\rangle$. One of the
mechanisms of decoherence of this superposition, that exists down
to zero temperature, is the decay of $|1\,\rangle$ onto
$|0\,\rangle$ accompanied by the radiation of the quantum of
energy $\Delta$.

As has been discussed in the Introduction, the conservation law
requires that the oscillations of the current are accompanied by
the torsional oscillations of the solid matrix, so that the total
angular momentum stays constant (e.g., zero). Such a local
wiggling of the matrix must result in the finite probability of
the emission of a transverse phonon of frequency $f_{0}$. Since
the wavelength of the phonon, ${\lambda}_s$, is large compared to
the dimensions of the SQUID, its effect on the SQUID is equivalent
to the uniform local rotation of the solid matrix at the position
of the SQUID \cite{Chu-Mar}. In terms of the deformation field
${\bf u}({\bf r},t)$, the angular velocity of this rotation is
given by
\begin{equation}\label{Omega}
{\bf \Omega}=\frac{1}{2}{\bf {\nabla}}{\times}\dot{{\bf u}}\;.
\end{equation}
Accordingly, the lattice velocity field at the position of the
SQUID is $\dot{\bf u}={\bf \Omega \times r}$. Substituting this
into Eq.(\ref{interaction}), one finds
\begin{equation}\label{rotational}
{\cal{H}}_{eff}={\hbar}{\bf L}{\cdot}{\bf \Omega}\;,
\end{equation}
where
\begin{equation}\label{j-L}
{\hbar}{\bf L} \, \equiv \, \frac{{\rm m}_e}{\rm e}\int\, d^3r\,\,
{\bf r} \times {\bf j}
\end{equation}
stands for the angular momentum of the SC current. The effective
interaction (\ref{rotational}), is mandated by symmetry and is,
therefore, parameter free. Correspondingly, the mechanism of
decoherence provided by this effect is universal.

Based upon Eq.\ (\ref{rotational}) the rate of the transition from
$|1\,\rangle$ to $|0\,\rangle$ is given by
\begin{equation}\label{CM-rate}
{\Gamma}_{s}=2{\hbar}<0|\hat{\bf
L}|1>[J_{env}({\Delta})]<1|\hat{\bf L}|0>\;,
\end{equation}
where $J_{env}(\Delta)$ is the spectral function of the
environmental coupling for phonons,
\begin{equation}\label{J-phonons}
J_{env}(\Delta)={\pi}\sum_{{\bf k},i}\langle{\bf k},i|\hat{\bf
\Omega}|0\rangle \langle0|\hat{\bf \Omega}|{\bf
k},i\rangle\delta(\Delta-{\hbar}{\omega}_{{\bf k}i})\;;
\end{equation}
${\bf k}$, $i$, and ${\omega}_{{\bf k}i}$ denote
the wave vector, polarization, and the frequency of the phonon,
respectively.
Further computation along the lines of Ref.\ \onlinecite{Chu-Mar}
yields
\begin{eqnarray}\label{CM-finalrate}
{\Gamma}_{s}(T) & = &
\frac{{\hbar}L^{2}}{12{\pi}{\rho}}\,k_{s}^{5}\coth\left[\frac{\Delta}{2k_{B}T}\right]
\\ \nonumber & = &
\frac{{\rm m}_{e}^{2}}{3{\pi}{\hbar}{\rm
e}^{2}}\,\frac{J^{2}a^{2}}{\rho}\,k_{s}^{5}\coth\left[\frac{\Delta}{2k_{B}T}\right]\;,
\end{eqnarray}
where  ${\rho}$ is the mass density of the solid matrix and
$k_{s}=2{\pi}/{\lambda}_{s}$ is the wave number of the emitted
sound.

It is important to notice that the decoherence rate of Eq.\
(\ref{CM-finalrate}) is proportional to the fifth power of $f_{0}$
and to the fourth power of the size of the SQUID. It is inversely
proportional to the fifth power of the speed of the transverse
sound, making it important to use solid matrices (substrates) of
high shear modulus. For practical values of the parameters:
$k_{B}T\,{\leq}\,\Delta$, $R\,{\sim}\,0.1\,{\mu}$m,
$\;J\,{\sim}\,0.1\,{\mu}$A, $\;{\rho}\,{\sim}\,5\,$g/cm$^{3}$,
$\;v_{s}\,{\sim}\,5{\times}10^{3}$m/s and
$f_{0}\,{\sim}\,5{\times}10^{9}\,$s$^{-1}$ (that is,
${\lambda}_{s}\,{\sim}\,1\,{\mu}$m), Eq.\ (\ref{CM-finalrate})
gives ${\Gamma}_{s}\,{\sim}\,1\,$s$^{-1}$. Consequently, the above
mechanism of decoherence should not be of great concern for a
small SQUID. For a large SQUID the situation will be quite
different, as is discussed below.

We should emphasize that Eq.\ (\ref{CM-finalrate}) can be used for
the estimate of the decoherence rate only at
$R\leq{\lambda}_{s}\,{\equiv}\,v_{s}/f_{0}$. In experiments of
Ref.\ \onlinecite{Friedman} and Ref.\ \onlinecite{Yu} the size the
SQUID was large in comparison with ${\lambda}_{s}$. The
interaction of such a SQUID with phonons is non-local and cannot
be treated by the above method. The non-local theory of the phonon
emission by a large SQUID is developed in the next Subsection.

\subsection{Large SQUID}

For $R\,>\,{\lambda}_{s}$, one should employ the general form of
the interaction given by Eq.\ (\ref{interaction}). This term
results in the coupled dynamics of the currents and the lattice
displacements. However, in reality, the large difference between
the ionic and electron masses makes the renormalization the SC
dynamics insignificant. In what follows, we will ignore the effect
of Eq.\ (\ref{interaction}) on the spatial and temporal structure
of the current and the SQUID flux $\Phi $ generated by the
current. We shall be concerned with the fact that, due to Eq.\
(\ref{interaction}), the currents serve as the source of phonons
in the elastic equation. In other words, the solid lattice must
take the recoil from the oscillating current. This effect is
mandated by conservation laws and it leads to the decoherence of
the quantum dynamics of the flux.

Inside a good conductor, either metal or a superconductor, the
longitudinal electric fields are screened with a typical time
scale of the plasma oscillations, $t\,{\sim}\,10^{-15}$s. The flux
dynamics is much slower. Consequently, the longitudinal phonons
that change the local concentration of ions should be excluded
from our consideration. This can be done by supplementing Eq.\
(\ref{interaction}) with the condition (\ref{transversal}). For
the purpose of estimates we shall adopt the simplest model of
uniform and isotropic elastic medium. Then, the energy of the free
transverse phonon field is
\begin{equation}\label{phonons}
{\cal{H}}_{ph}=\int\,d^{3}r \left(\frac{1}{2}{\rho}{\dot{\bf
u}}^{2}+ {\mu}u_{ij}^{2}\right)\;,
\end{equation}
where ${\mu}$ is the shear modulus of the solid and
$u_{ij}=\frac{1}{2}({\partial}_{i} u_{j}+{\partial}_{j}u_{i})$ is
the strain tensor ($\sum_{i}u_{ii}$ being zero for transverse
phonons). In this model the torsional strains are described by
just one elastic modulus ${\mu}={\rho}v_{s}^{2}$. Accordingly, the
transverse sound velocity $v_s$ is independent of the phonon
polarization (which is orthogonal to the phonon wave vector). We
shall further simplify our consideration by neglecting all
differences in the actual material composition of the experimental
setup, that is, by assuming that the phonon spectrum is the same
inside and outside the part of the solid matrix that carries the
SC current. This assumption, while not valid in experiment, should
not significantly affect our estimate of the decoherence rate.

The canonical quantization of the phonon field yields:
\begin{eqnarray}\label{quantization}
{\bf u}({\bf r}) & = & \frac{1}{\sqrt{V}} \sum_{{\bf k},i}
\sqrt{\frac{\hbar}{2{\rho}{\omega}_{{\bf k}i}}}\left( a_{{\bf
k}i}{\rm e}^{i\bf k r} + a^\dagger_{{\bf k}i}{\rm e}^{-i\bf k r}
\right){\bf e}_{i} \nonumber \\
{\cal{\bf \Pi}}({\bf r}) & = &  \frac{-i}{\sqrt{V}} \sum_{{\bf
k},i}\sqrt{\frac{{\hbar}{\omega}_{{\bf k}i}{\rho}}{2}}\left(
a_{{\bf k}i}{\rm e}^{i\bf k r} - a^\dagger_{{\bf k}i}{\rm
e}^{-i\bf k r} \right){\bf e}_{i}\;. \nonumber \\
& &
\end{eqnarray}
Here ${\cal{\bf \Pi}} = {\rho}{\dot{\bf u}}$ is the momentum of
the phonons that is canonically conjugate to ${\bf u}$,
$\;{\omega}_{{\bf k}i}=v_{s}k$ is the frequency of the phonon of
the wave vector ${\bf k}$ and polarization $i$, and $V$ is the
volume of the system. Due to the isotropy, $\omega_{{\bf k}i}$ for
the transverse phonons (${\bf k}{\cdot}{\bf e}_i=0$) does not
depend on the polarization. Substituting Eqs.\
(\ref{quantization}) into Eq.\ (\ref{phonons}) and Eq.\
(\ref{interaction}) one obtains
\begin{eqnarray}\label{total}
{\cal{H}} & = & {\cal{H}}_{ph}+{\cal{H}}_{int} = \sum_{{\bf
k},i}\hbar \omega_{{\bf k}i}\left(a^\dagger_{{\bf k}i} a_{{\bf
k}i} +
\frac{1}{2}\right) \nonumber \\
& + & \frac{i}{\sqrt{V}}\sum_{{\bf k},i} \, \frac{{\rm m}_{e}}{\rm
e}\left(\frac{\hbar \omega_{i\bf k}}{2{\rho}}\right)^{1/2}({\bf
j}_{\bf k}{\cdot}{\bf e}_{i})( a_{{\bf k}i} - a^{\dagger}_{{\bf
k}i})\;, \nonumber \\
& &
\end{eqnarray}
where ${\bf j}_{\bf k}= \int d^3r\, {\bf j}\,\,\exp (-i{\bf kr})$
is the spatial Fourier component of the current density ${\bf
j}({\bf r})$, and the summation is over ${\bf k}$ and $i$
satisfying ${\bf k}{\cdot}{\bf e}_{i}=0$. The Fermi golden rule,
then, yields the following expression for the decoherence rate:
\begin{eqnarray}\label{rate}
\Gamma_{ph} & = & \frac{2\pi}{\hbar} \frac{{\rm m}^2_e}{2{\rm
e}^2{\rho}}\, \Delta \coth
\left[\frac{\Delta}{2k_{B}T}\right]{\times} \nonumber
\\ & &
\sum_{i}\int  \frac{d^3k}{ (2\pi )^3} |\langle\,0\,|({\bf j}_{\bf
k}{\cdot}{\bf e}_{i})|\,1\,\rangle|^2 \delta (\Delta - \hbar
\omega_{i\bf k})\;, \nonumber \\
& &
\end{eqnarray}
where $|\,0\,\rangle$ and $|\,1\,\rangle$ are given by Eqs.\
(\ref{ground}) and (\ref{first}).

We shall now compute ${\bf j}_{\bf k}$. The SC current can be
written in terms of the SC phase $\varphi$ and the vector
potential $\bf A$,
\begin{equation}\label{current}
{\bf j}= \frac{c}{ 4\pi \lambda_L^2} \left(\frac{\Phi_0 }{ 2\pi}
{\bf {\nabla}} \varphi - {\bf A}\right)\;,
\end{equation}
with $\Phi_0 = hc/ 2{\rm e}$ and $\lambda_L$ being the flux
quantum and the London penetration length, respectively. The
vector potential satisfies the Maxwell equation,
\begin{equation}\label{MAX}
{\bf {\nabla}} \times {\bf {\nabla}} \times {\bf A}=
\frac{4{\pi}}{c}{\bf j}\;,
\end{equation}
with the current density ${\bf j}$ given by Eq.\ (\ref{current})
inside the SC loop and by ${\bf j}=0$ outside the loop.

To simplify calculations we shall adopt the ring geometry of the
SQUID. In the presence of the flux $\Phi$, the phase $\varphi$
winds around the ring by ${\varphi}_J = 2\pi \Phi /\Phi_0$, which
is the Josephson phase in the junction cutting the ring. We shall
study the problem in cylindrical coordinates ($z,r,\phi$), with
the $Z$-axis passing through the center of the ring perpendicular
to its plane, $r$ standing for the radial coordinate, and $\phi$
being the polar angle in the plane of the ring. The solution for
the phase is
\begin{equation}\label{phase}
\varphi = \frac{\varphi_J \phi }{ 2\pi}= \frac{\Phi }{
\Phi_0}\phi\;.
\end{equation}
Due to the cylindrical symmetry of the currents, the only non-zero
component of the vector potential is $A\equiv A_\phi (r, z)$.
Then, inside the ring, Eq.\ (\ref{MAX}) reduces to
\begin{equation}\label{vector}
-\frac{\partial }{ \partial r}\left[\frac{1}{r} \frac{\partial }{
\partial r}(r A)\right] - \frac{{\partial^2}A}{
\partial^2 z} +\lambda^{-2}_LA=\frac{\Phi }{ 2\pi \lambda^2_L}
\frac{1}{ r}\;,
\end{equation}
while outside the ring one has
\begin{equation}\label{vector_out}
-\frac{\partial }{ \partial r}\left[\frac{1}{ r} \frac{\partial }{
\partial r}(r A)\right] - \frac{{\partial^2}A}{
\partial^2 z} = 0\;.
\end{equation}
These equations must be accompanied by the boundary conditions for
$A$ and for non-zero components of the magnetic field,
\begin{equation}\label{fields}
H_r = -\frac{\partial A}{\partial z}\;,\;\;\;\; H_z=\frac{1}{r}
\frac{\partial (r A)}{\partial r}\;.
\end{equation}
In this paper we shall not pursue the exact solution of the
problem for the finite cross-section of the ring carrying the
current. Instead, we will make use of a thin-ring approximation in
which the thickness of the ring $\sqrt{b}$,
where $b$ stands for the area of the wire
crossection, is small compared to its radius
$R$ as well as to $\lambda_L$. Then, in cylindrical coordinates,
the only non-zero component of the current density is
$j_\phi(z,r)$. It equals $J/b$ inside the ring and zero outside
the ring. At
$k \sqrt{b} \ll 1$ the Fourier transform of such a distribution of
the current is
\begin{eqnarray}\label{j_found}
{\bf j}_{\bf k}= - i 2{\pi}R J_1(k_{\perp}R)J{\bf n}_{\bf k}\;,
\end{eqnarray}
where ${\bf n}_{\bf k}\,{\perp}\,\bf k$ is the unit vector in the
plane of the ring, $k_{\perp}=k\sin\theta$, $\,\theta$ is the
angle between $\bf k$ and the $Z$-axis, and $J_1(k_{\perp}R)$
stands for the Bessel function of the first order.

The quantization procedure consists of assigning the operator
$\hat{J}$ to the total current $J$. In the two-level
approximation, one introduces the states $|\pm \rangle$ of the
current operator such that $\pm J$ are the respective eigenvalues:
$\hat{J}|\pm \rangle = \pm J|\pm \rangle$. In terms of the angular
momentum operator, these states are identical to those in Eq.\
(\ref{L_z}), that is $|  +  \rangle  \equiv  | \uparrow \,
\rangle$ and  $|  -  \rangle  \equiv  | \downarrow \, \rangle$.
Tunneling between these two degenerate states produces new states,
Eq.\ (\ref{ground}) and Eq.\ (\ref{first}), which are split by the
energy $\Delta$. These states are characterized by the zero
current, $\langle\,0\,|\hat{J}|\,0\,\rangle = 0$ and
$\langle\,1\,|\hat{J}|\,1\,\rangle = 0$. The transition matrix
element is $\langle\,0\,|\hat{J}|\,1\,\rangle = J$. Substituting
Eq.\ (\ref{j_found}) into Eq.\ (\ref{rate}) we get
\begin{eqnarray}\label{thin}
\Gamma_{ph} & = & \frac{2{\pi}{{\rm m}_e}^2}{{\hbar}{\rm
e}^{2}}\frac{J^{2}R^2}{\rho}\, \coth
\left[\frac{\Delta}{2k_{B}T}\right]{\times} \nonumber \\
& &  k^3_s\int^1_0 d\cos\theta \; J_1^2(k_sR\sin \theta) .
\end{eqnarray}
The limit of $k_s R \ll 1$
corresponds to a small SQUID. It is easy to see that in this limit
${\Gamma}_{ph}$ is proportional to $k_s^5$ and Eq.\ (\ref{thin})
becomes Eq.\ (\ref{CM-finalrate}). In the limit of a large SQUID,
$k_s R\gg 1$, we find
\begin{equation}\label{kR>>1}
\Gamma_{ph}=\frac{{\pi} {\rm m}_e^2}{{\hbar}{\rm e}^{2}}
\frac{J^{2}R}{\rho}\,k^2_s
\coth\left[\frac{\Delta}{2k_{B}T}\right].
\end{equation}

Based upon Eq.\ (\ref{kR>>1}), let us make an estimate of the
decoherence rate for, e.g., the experiment of Ref.\
\onlinecite{Friedman}. At $k_{B}T\,{\leq}\,\Delta$, for
$R\,{\sim}\,0.1\,$mm, $\;J\,{\sim}\,3\,{\mu}$A,
$\;{\rho}\,{\sim}\,8\,$g/cm$^{3}$,
$\;v_{s}\,{\sim}\,5{\times}10^{3}$m/s and
$f_{0}\,{\sim}\,2{\times}10^{9}\,$s$^{-1}$ (that is,
${\lambda}_{s}\,{\sim}\,2.5\,{\mu}$m), Eq.\ (\ref{kR>>1}) gives
${\Gamma}_{ph}\,{\sim}\,10^{6}\,$s$^{-1}$. This is a significant
decoherence rate that would limit the quality factor of the
corresponding qubit by the value of about one thousand.

\subsection{Global noise}
Here we will compare the effect of the global noise ---
uncontrolled rotations of the solid matrix as a whole at some
angular velocity ${\Omega}_{G}(t)$ --- with the above estimates
for the local effects due to phonons. The $Z$-component of the
global rotation removes the degeneracy between clockwise and
counterclockwise current states. This is a particular case of the
Barnett effect: A rotating solid develops magnetization
proportional to the angular velocity of the rotation
\cite{Barnett}. In application to the SQUID this effect can be
described by a two-state Hamiltonian written in the rotating
coordinate frame:
\begin{equation}\label{global}
{\cal{H}}_{G}=-{\Delta}s_{x} - 2{\hbar}L{\Omega}_{G}(t)s_{z}\;,
\end{equation}
where $s_{x,z}$ are spin-1/2 operators. We want to estimate the
effect of the second term in Eq.\ (\ref{global}) on coherent
oscillations of the SC current. (Notice that for externally
imposed rotations, the sign of this term is opposite to the sign
of Eq.\ (\ref{rotational}) that was written for phonons
dynamically produced by SQUID oscillations in the laboratory
frame.) For a macroscopic solid matrix, the characteristic
correlation time of $\langle {\Omega}_{G}(t) {\Omega}_{G}(0)
\rangle$ cannot be less than the time it takes the sound to travel
across the matrix. Consequently, on the time scale
$t\,{\sim}\,1/f_0$, random rotations of the equipment as a whole
are slow enough to permit the treatment of Eq.\ (\ref{global})
within the adiabatic approximation.

In the adiabatic approximation the eigenvalues and eigenfunctions
of the Hamiltonian (\ref{global}) are
\begin{eqnarray}\label{eigen}
{\epsilon}_{\pm} & = & \pm\sqrt{{\Delta}^{2} +
[2{\hbar}L{\Omega}_{G}]^{2}}
\nonumber \\
{\psi}_{\pm} & = & \frac{1}{\sqrt{2}}\,[C_{\mp}(t)\,|\uparrow \,
\rangle \; {\pm} \; C_{\pm}\,|\downarrow \, \rangle] \;,
\end{eqnarray}
where $C_{\pm}$ are given by
\begin{equation}\label{C-coefficients}
C_{\pm}=\sqrt{1 \, \pm \,
\frac{2{\hbar}L{\Omega}_{G}}{\sqrt{{\Delta}^{2}+[2{\hbar}L{\Omega}_{G}]^{2}}}}
\;.
\end{equation}
One can see that random rotations of the system as a whole do not
significantly perturb the states $|\,0\,\rangle$ and
$|\,1\,\rangle$ given by Eq.\ (\ref{ground}) and Eq.\
(\ref{first}) only if ${\Omega}_{G}$ satisfies
\begin{equation}\label{Omega-G}
{\Omega}_{G}\,\ll\,\frac{{\omega}_{0}}{2L}\;.
\end{equation}
For a large SQUID with $L\,{\sim}\,10^{10}$ and
${\omega}_{0}\,{\sim}\,10^{10}\,$s$^{-1}$, this gives
${\Omega}_{G} \ll 1\,$s$^{-1}$, which must be of practical
importance in the situation when the equipment is subjected to
random movements. For a small SQUID, with small $L$ and practical
values of ${\omega}_{0}$, the condition (\ref{Omega-G}) can be
satisfied by a very large margin.

The time dependence of $\Omega_G$ is another factor. The
adiabaticity implies that
\begin{equation}\label{adiab}
{\cal A}\equiv \frac{{\pi}\omega_0^2}{4L|\dot{\Omega}_{G}|}\gg
1\;.
\end{equation}
Then, the main effect of the time dependence of the global
rotations is generation of additional harmonics in the Rabi
oscillations between clockwise and counterclockwise SC currents
due to the time dependence of ${\epsilon}_{\pm}$. The
corresponding decoherence rate, $\Gamma_G$, can be estimated from
the variation of ${\epsilon}_{\pm}$ during one cycle,
$2{\pi}/{\omega}_{0}$, of the undisturbed Rabi oscillations. The
global random rotations occur when the equipment is subjected to a
random external torque. Let this torque result in an angular
acceleration ${\alpha}=\dot{\Omega}_G$, so that the change of
$\Omega_G$ during one cycle is $\delta {\Omega}_{G}\,{\sim}\,
{\alpha}/\omega_0$. This gives
\begin{equation}\label{Gamma_G}
{\Gamma}_{G}\,{\sim}\,
\delta{\epsilon}_{+}/{\hbar}\,{\sim}\,{\omega}_{0}/{\cal A}\;,
\end{equation}
where ${\cal A}= {\pi}\omega_0^2/4L|\alpha|$. Thus, the quality
factor for the above mechanism is $Q=\cal{A}$. It is entirely
determined by the adiabaticity of global rotations.

If the typical frequency of the mechanical noise is
${\omega}_{n}$, then ${\alpha}\,{\sim}\,{\omega}_{n}{\Omega}_{G}$.
Assuming that the condition (\ref{Omega-G}) is satisfied, this
implies
\begin{equation}\label{Q-A}
Q = {\cal A} \gg {\omega}_{0}/{\omega}_{n}\;.
\end{equation}
In principle, at small ${\omega}_{0}$, the effect of the
mechanical noise can overpower the decoherence from local phonon
effects, which decrease as some power of $\omega_0$. However, for
any practical values of ${\omega}_{0}$ and $\alpha$ in the
coherence experiments, the adiabadicity factor ${\cal A}$ is very
large, so that the decohering effect of the external global noise
can be safely ignored.

We also would like to make the following interesting observation.
Consider rotations ${\Omega}_{G}={\alpha}t$ that last long enough
to violate the condition (\ref{Omega-G}). According to Eq.\
(\ref{C-coefficients}), as time goes from $-\infty$ to $+\infty$,
the states ${\psi}_{\pm}$ of Eq.\ (\ref{eigen}) switch between
$|\uparrow \, \rangle$ and $|\downarrow \, \rangle$. This,
however, is true only in the limit of
$|{\alpha}|\,{\rightarrow}\,0$. At finite $|\alpha|$ the answer
depends on the adiabaticity factor ${\cal A}$. If at $t=-\infty$
the SQUID is prepared in, e.g., the state with the clockwise SC
current, then, the probability for the SQUID to switch to the
counterclockwise current at $t=+\infty$ is given by the
Landau-Zener formula \cite{LZ}
\begin{equation}\label{LZ}
P_{LZ}=1-e^{-{\cal A}}\;.
\end{equation}
This probability  is high if the angular acceleration satisfies
${\alpha}\,\leq\,{\alpha}_{c}={\pi}{\omega}_{0}^{2}/4L$.
Consequently, at low tunneling rate the relatively slow mechanical
rotation can provide the quantum-mechanical switching between
clockwise and counterclockwise currents. It should be noted,
however, that the Landau-Zener transitions can also be generated
by the magnetic field. Thus, the above effects of uniform
rotations can only be observed if the SQUID is shielded from the
magnetic fields with an accuracy
${\mu}_{B}H\,<\,{\hbar}{\Omega}_{G}$.

\section{Decoherence due to photons}

The problem of decoherence due to the emission of photons of
frequency $f_0$ is very similar to the problem of the emission of
phonons. The main difference is that the vacuum wavelength of the
light, ${\lambda}_{l}=c/f_{0}$, is typically large compared to the
size of the SQUID that exhibits quantum oscillations of the
current. The electromagnetic radiation by such a SQUID into the
open space is equivalent to the radiation of a point magnetic
dipole. If, however, the SQUID is shielded by a metal placed at a
distance that is comparable to or smaller than ${\lambda}_{l}$,
the decoherence rate becomes strongly geometry-dependent. These
two problems are considered in the following two Subsections.

\subsection{Decoherence in the open space}

If the wavelength of the emitted photons is large compared with
the SQUID size, then the SQUID can be treated as a point particle
with an angular momentum ${\bf L}$ which is perpendicular to the
SQUID loop. This angular momentum interacts with the photon field
via Zeeman Hamiltonian
\begin{equation}\label{Zeeman}
{\cal{H}}_{Z}=-{\mu}_{B}L_{z}H_{z}\;\;,
\end{equation}
where $H_z$ is the $Z$-component of the magnetic field of the
vacuum photons. We are interested in the transition between the
tunnel-splitted quantum states given by Eq.\ (\ref{ground}) and
Eq.\ (\ref{first}). The expression for the rate is similar to Eq.\
(\ref{CM-rate}), where the spectral density of the photons can be
obtained by either quantizing ${\bf H}({\bf r})$ or taking
$\langle\,|H_{\omega}|^{2}\rangle$ from the theory of
electromagnetic fluctuations \cite{Chu-Gar,Landau}.

The Fermi golden rule then yields the following expression for the
decoherence rate
\begin{eqnarray}\label{Gamma-Fermi}
{\Gamma}_{Z}(T) & = &
\frac{4}{3\hbar}\,{\mu}_{B}^{2}L^{2}k_{l}^{3}\coth\left[\frac{\Delta}{2k_{B}T}\right]
\\ \nonumber
& = &
\frac{4}{3{\hbar}c^{2}}\,J^{2}a^{2}k_{l}^{3}\coth\left[\frac{\Delta}{2k_{B}T}\right]\;,
\end{eqnarray}
where $k_{l}={\Delta}/{\hbar}c$ is the wave vector of the emitted
light. It is proportional to the third power of $f_{0}$ as
compared to the fifth power of $f_{0}$ in Eq.\
(\ref{CM-finalrate}). The reason for the difference is the
additional time derivative of the boson field in the Hamiltonian
of Eq.\ (\ref{rotational}) as compared to Eq.\ (\ref{Zeeman}).

Notice that the identical result for ${\Gamma}_{Z}$ follows from
\begin{equation}\label{classical}
{\Gamma}_{Z}(0)=\frac{I}{{\hbar}{\omega}_{0}}\;,
\end{equation}
where \cite{Landau}
\begin{equation}\label{intensity}
I=\frac{4}{3c^{3}}|\ddot{\bf M}|^{2}
\end{equation}
is the intensity of the magnetic dipole radiation due to the
classical dynamics of the magnetic moment ${\bf M}={\mu}_{B}{\bf
L}$.

For a small SQUID (as defined above in Sec. II)  with
$\;R\,{\sim}\,{\lambda}_{s}\,{\sim}\,100\,$nm and
$J\,{\sim}\,0.1\,{\mu}$A, Eq.\ (\ref{Gamma-Fermi}), at
$f_{0}\,{\sim}\,10^{10}\,$s$^{-1}$ and $k_{B}T \leq \Delta$, gives
negligible decoherence, ${\Gamma}_{Z}\,{\sim}\,10^{-8}\,$s$^{-1}$.
However, for a large SQUID with $R\,{\sim}\,100\,{\mu}$m and
$J\,{\sim}\,3\,{\mu}$A, at $f_{0}\,{\sim}\,10^{10}\,$s$^{-1}$ and
$k_{B}T \ll \Delta$, one obtains
${\Gamma}_{Z}\,{\sim}\,10^{7}\,$s$^{-1}$. Thus, for a large SQUID,
the radiation of photons into the open space can easily reduce the
quality factor of the SQUID down to one hundred.

The decoherence rate can be decreased by choosing a double-loop
geometry with equal areas of the single loops and equal currents
flowing in the opposite directions, as was actually done in Ref.
\onlinecite{Friedman}. In that case, the total magnetic moment of
the system is zero and the radiation is of the quadrupolar nature.
If the magnetic moment is compensated exactly (which must be
difficult to achieve in experiment) the radiation rate will be
reduced by a factor $(kR)^{2} \ll 1$. At a finite compensation,
$\gamma={\Delta}M/M\,<\,1$, the decoherence rate
(\ref{Gamma-Fermi}) acquires a factor ${\gamma}^{2}$.

Here we have neglected the effects of dc and low-frequency
adiabatic ac magnetic fields $H(t)$. These effects are equivalent
to the effects of global rotations studied in the Subsection II-C.
To estimate them quantitatively one should replace ${\Omega}_{G}$
and ${\alpha}$ by ${\omega}_{H}={\rm e}H/2m_{e}c$ and
${\omega}_{ac}{\omega}_{H}$ respectively, where ${\omega}_{ac}$ is
the typical frequency of the ac field. Then, conditions
(\ref{Omega-G}) and (\ref{adiab}) become
\begin{equation}\label{omega-H}
{\omega}_{H}\,\ll\,\frac{{\omega}_{0}}{2L}\;,
\end{equation}
and
\begin{equation}\label{adiab-H}
{\cal A}\equiv
\frac{{\pi}\omega_0^2}{4L{\omega}_{ac}{\omega}_{H}}\gg 1\;,
\end{equation}
respectively. In a two-loop design, where the magnetic moment is
compensated by a factor $\gamma={\Delta}M/M\,<\,1$, $\;L$ in Eqs.\
(\ref{omega-H}) and (\ref{adiab-H}) should be replaced by
${\gamma}L$.

For, e.g, a small SQUID with $L\,{\sim}\,100$ and no compensation,
Eq.\ (\ref{omega-H}) at ${\omega}_{0}\,{\sim}\,10^{10}$s$^{-1}$
translates into $H \ll 10\,$Oe. For a large SQUID with
${\gamma}=10^{-2}$, $\;L=10^{10}$ and
${\omega}_{0}\,{\sim}\,10^{10}$s$^{-1}$, the fields that do not
disturb the states $|\,0\,\rangle$ and $|\,1\,\rangle$ should
satisfy $H \ll 10^{-5}\,$Oe. If Eq.\ (\ref{omega-H}) is satisfied,
then substituting it into Eq.\ (\ref{adiab-H}) one obtains the
following relation for the quality factor coming from the
low-frequency magnetic noise alone:
\begin{equation}\label{Q}
Q = {\cal A} \gg {\omega}_{0}/{\omega}_{ac}\;.
\end{equation}

\subsection{Decoherence in the presence of metal shielding}

We shall now study the case when the SQUID loop is adjacent to a
metal sheet parallel to the plane of the loop. Following published
experiments, we shall assume that the distance between the SQUID
and the shielding, $d$, is much smaller than the wavelength of the
vacuum electromagnetic radiation ${\lambda}_{l}$. In that case the
radiation becomes strongly renormalized by the conducting medium
and the formulas of the previous Subsection can no longer be used.
Now the main source of decoherence is the dissipative current in
the metal shielding induced by the ac fields of the SQUID.
Correspondingly, the decoherence rate can be computed as
\begin{equation}\label{Gamma-M}
{\Gamma}_{M}=\frac{P}{{\hbar}{\omega}_{0}}\;,
\end{equation}
where $P$ is the power absorbed by the shielding. Let ${\bf
H}={\bf H}_{\omega}({\bf r})\exp(i{\omega}_{0}t)$ be the magnetic
field generated by the oscillating current in the SQUID. Then $P$
is given by \cite{Landau}
\begin{equation}\label{power}
P=\frac{c}{16\pi}\sqrt\frac{{\omega}_{0}}{2{\pi}{\sigma}}\oint{\langle}\,|{\bf
H}_{\omega}|^{2}{\rangle}df\;,
\end{equation}
where $\sigma$ is the electric conductivity of the shielding,
$\langle...\rangle$ means quantum-mechanical average, and the
integration goes over the metal surface facing the SQUID.

Eq.\ (\ref{power}) can be used when the thickness of the skin
layer, ${\delta}=c/\sqrt{2{\pi}{\sigma}{\omega}_{0}}$, is small
compared to the thickness of the shielding metal, $D$, but large
compared to the mean free path of electrons of the metal, $l_{0}$.
These conditions were apparently fulfilled in the experiment of
Ref.\ \onlinecite{Friedman} for which we estimate
${\delta}\,{\sim}\,1\,{\mu}$m and $l_{0}<0.1\,{\mu}$m at
$D\,{\sim}\,8\,{\mu}$m. The condition $d \ll {\lambda}_{l}$ allows
one to use the quasistationary approximation \cite{Landau} to
obtain ${\bf H}$. In this approximation the field is formed by two
current loops, one being the mirror image of the other with
respect to the surface of the shielding. At the metal surface,
$z=d$, this field has the tangential component only, $H_r(r)$. For
a thin circular loop carrying the electric current $J(t)=J \,
\exp(i{\omega}_{0}t)$, it is given by
\begin{equation}\label{rho-component}
H_r=\frac{4 J R d}{c} \int_0^\pi \frac{ d\phi \, \cos \phi }{
(d^2+r^2+ R^2 - 2r R\cos{\phi})^{3/2}}\;,
\end{equation}
which can be expressed in terms of the elliptic integrals
\cite{Landau}.

Substituting Eq.\ (\ref{rho-component}) into Eq.\ (\ref{power}),
one obtains
\begin{equation}\label{Gamma-shield}
{\Gamma}_{M}=\frac{2}{hc}\frac{J^{2}}{\sqrt{{\sigma}f_{0}}}\,
F(R/d)\;,
\end{equation}
where the
function $F$ is given by
\begin{equation}\label{F}
F(\xi) = \xi^2\int_{0}^{\infty}dx\,x
\left[\int_{0}^{\pi}\frac{\cos{\phi}\,d{\phi}}{(1+x^{2}+{\xi}^{2}
-2x{\xi}\cos{\phi})^{3/2}}\right]^{2}\;.
\end{equation}
In the two limiting cases of small and large $d$, the geometrical
factor (\ref{F})  reduces to
\begin{eqnarray}\label{F-limits}
F & = & \frac{l}{4d}\;,\;\;\;\;d \ll R \\ \nonumber F & = &
\frac{3a^{2}}{32d^{4}}\;,\;\;\;\;d \gg R\;.
\end{eqnarray}
Here we have introduced the length and the area of the current
loop, $l=2{\pi}R$ and $a={\pi}R^{2}$, respectively, in order to
emphasize the fact that the above limiting expressions are correct
for flat current loops of arbitrary shape. Numerical analysis
shows that these expressions hold for $d/R \leq 0.3$ and $d/R
>5$, correspondingly. For the double-loop with a compensated magnetic
moment due to equal single-loop currents flowing in the opposite
directions, the decoherence rate practically does not change in
the limit of $d \ll R$. In the opposite limit of $d \gg R$, it
reduces by a factor $(R/d)^{2}$.

Note that the frequency dependence of the rate
(\ref{Gamma-shield}) follows from the frequency dependence of the
skin depth, $\delta \propto  1/\sqrt{f_0}$. Then the Maxwell
equation gives $E \propto \delta {\cdot} (\partial H/\partial t)
\propto \sqrt{f_0}$ for the electric field in the skin layer and
the dissipation rate due to Joule's power losses, $\sigma E^2
\delta$, becomes proportional to $\sqrt{f_0}$. Divided by $f_{0}$
in Eq.\ (\ref{Gamma-M}), it gives the $1/\sqrt{f_0}$ dependence of
the decoherence rate. If the thickness of the metal shielding $D$
is smaller than $\delta$, the electric and magnetic fields are not
significantly modified by the shielding, so that $E \propto
f_{0}H$, with the proportionality factor determined by the
geometry of the SQUID and its distance to the shielding. Thus, in
the low-frequency limit ($D\,<\,\delta$), the dissipation in the
shielding $P \,{\sim} \, \sigma E^2 D$ is proportional to $f_0^2$
and $\Gamma_M$ due to shielding is proportional to $f_0$.

For $J\,{\sim}\,3\,{\mu}$A, $\;{\sigma} = 3{\times}10^{17}$e.m.u.,
$\;f_{0}\,{\sim}\,2{\times}\,10^{9}$s$^{-1}$, and $R\,{\sim}\,d$,
Eq.\ (\ref{Gamma-shield}) gives
${\Gamma}_{M}\,{\sim}\,10^{9}$s$^{-1}$, which was, probably, the
case in the experiment of Ref.\ \onlinecite{Friedman}. This shows
that for a SQUID carrying a microampere current the above
mechanism can provide a very high decoherence rate. Notice that
${\Gamma}_{M}$ can be drastically reduced by increasing the
distance, $d$, between the SQUID and the shielding. Indeed,
according to Eqs.\ (\ref{F-limits}),
${\Gamma}_{M}\,{\propto}\;d^{-4}$ at $R \ll d \ll {\lambda}_{l}$.
Too large $d$, however, would reduce the effectiveness of the
shielding in protecting the SQUID from external radio signals.
Choosing smaller SQUIDS operating at smaller currents should be
more beneficial for qubit
designs.\\


\section{Conclusions}
We have studied generic mechanisms of decoherence mandated by the
conservation laws --- emission of phonons and photons of the
oscillation frequency $f_{0}$. Our practical conclusions are as
follows:

The decoherence due to the above mechanisms scales as the second
power of the current.

For small SQUIDS of size $R\,<\, v_{s}/f_{0}$, the decoherence due
to the emission of phonons at $T \leq {\Delta}$ is negligible.

For large SQUIDs of size $R \gg v_{s}/f_{0}$, the emission of
phonons can significantly limit the quality factor. The
corresponding decoherence rate scales linearly with the size of
the SQUID and quadratically on the oscillation frequency.

In the absence of the metal shielding, the emission of photons is
negligible for small SQUIDs but becomes significant for large
SQUIDs. It scales as the fourth power of the size of the SQUID and
as the third power of the oscillation frequency.

Decoherence due to the shielding strongly depends on the geometry
of the experimental setup. It may completely destroy the coherence
in large SQUIDs and can be the main mechanism of decoherence in
small SQUIDs. The shielding must be provided by a metal sheet of
thickness greater than the skin layer at the oscillation frequency
$f_{0}$. To achieve small decoherence, the distance to the
shielding, while small in comparison with $c/f_{0}$, should be
considerably greater than the loop size.

The effects of the external mechanical and magnetic noise are
proportional to the total magnetic moment of the SQUID, making
small SQUIDS less susceptible to the noise than large SQUIDs.
\\

\section*{Acknowledgments}
We thank J. Friedman and B. Svistunov for useful discussions. This
work has been supported by the DOE Grant No. DE-FG-2-93ER45487.


\begin{thebibliography}{10}

\bibitem{Cal-Leg}
{A. O. Caldeira and A. J. Leggett}, Ann. Phys. (N.Y.) {\bf 149},
374 (1983).


\bibitem{Martinis}
{J. M. Martinis, M. H. Devoret, and J. Clarke}, Phys. Rev. Lett.
{\bf 55}, 1543 (1985).


\bibitem{Clarke}
{J. Clarke, A. N. Cleland, M. H. Devoret, D. Esteve, and J. M.
Martinis}, Science {\bf 239}, 992 (1988).

\bibitem{Leggett}
{A. J. Leggett, S. Chakravarty, A. T. Dorsey, M. P. Fisher, A.
Garg, W. Zwerger}, Rev. Mod. Phys. {bf 59}, 1 (1987).

\bibitem{Friedman}
{J. R. Friedman, V. Patel, W. Chen, S. K. Tolpygo, and J. E.
Lukens}, Nature {\bf 406}, 43 (2000).

\bibitem{Mooij}
{C. H. van der Wal, A. C. J. ter Haar, F. K. Wilhelm, R. N.
Schouten, C. J. P. M. Harmans, T. P. Orlando, S. Lloyd, and J. E.
Mooij}, Science {\bf 290}, 773 (2000).

\bibitem{Vion}
{D. Vion, A. Aassime, A. Cottet, P. Joyez, H. Pothier, C. Urbina,
D. Esteve, M. Devoret}, Science {\bf 296}, 886 (2002).

\bibitem{Yu}
{Y. Yu, S. Han, X. Chu, S. Chu, Z. Wang}, Science {\bf 296}, 889
(2002).

\bibitem{Leggett-Science}
See, e.g., {A. J. Leggett}, Science {\bf 296}, 861 (2002).

\bibitem{Chudnovsky}
{E. M. Chudnovsky}, Phys. Rev. Lett. {\bf 72}, 3433 (1994).

\bibitem{Chu-Mar}
{E. M. Chudnovsky and X. Martinez-Hidalgo}, Phys. Rev. {\bf B66},
054412 (2002).

\bibitem{Barnett}
{S. J. Barnett}, Rev. Mod. Phys. {\bf 7}, 129 (1937).

\bibitem{LZ}
{L. D. Landau}, Phys. Z. Sowjetunion {\bf 2}, 46 (1932); {C.
Zener}, Proc. R. Soc. London, Ser. A, {bf 137}, 696 (1932).

\bibitem{Chu-Gar}
{E. M. Chudnovsky and D. A. Garanin}, Phys. Rev. Lett. {\bf 89},
157201 (2002).

\bibitem{Landau}
{L. D. Landau and E. M. Lifshitz}, {\it Electrodynamics of
Continuous Media} (Nauka,Moscow, 1982).

\end{thebibliography}

\end{document}